\documentclass[aps,twocolumn,superscriptaddress]{revtex4-2}

\usepackage{amsmath, amssymb, graphicx, subfigure}

\begin{document}

\title{Robust anomalous metallic states and vestiges of self duality in two-dimensional granular In-InOx composites}

\author{Xinyang Zhang}
\affiliation{Geballe Laboratory for Advanced Materials, Stanford University, Stanford, CA 94305}
\affiliation{Department of Applied Physics, Stanford University, Stanford, CA 94305}

\author{Bar Hen}
\affiliation{School of Physics and Astronomy, Raymond and Beverly Sackler,\\
Faculty of Exact Sciences, Tel Aviv University, Tel Aviv 6997801, Israel}

\author{Alexander Palevski}
\affiliation{School of Physics and Astronomy, Raymond and Beverly Sackler,\\
Faculty of Exact Sciences, Tel Aviv University, Tel Aviv 6997801, Israel}

\author{Aharon Kapitulnik}
\affiliation{Geballe Laboratory for Advanced Materials, Stanford University, Stanford, CA 94305}
\affiliation{Department of Applied Physics, Stanford University, Stanford, CA 94305}
\affiliation{Department of Physics, Stanford University, Stanford, CA 94305}

\date{\today}

\begin{abstract}
Many experiments investigating magnetic field tuned superconductor-insulator transition (H-SIT) often exhibit low-temperature resistance saturation, which is interpreted as an anomalous metallic phase emerging from a ``failed superconductor," thus challenging conventional theory. Here we study a random granular array of indium islands grown on a gateable layer of indium-oxide. By tuning the intergrain couplings, we reveal a wide range of magnetic fields where resistance saturation is observed, under conditions of careful electromagnetic filtering and within a wide range of linear response. Exposure to external broadband noise or microwave radiation is shown to strengthen the tendency of superconductivity, where at low field a global superconducting phase is restored. Increasing magnetic field unveils an ``avoided H-SIT'' that exhibits granularity-induced logarithmic divergence of the resistance/conductance above/below that transition, pointing to possible vestiges of the original emergent duality observed in a true H-SIT. We conclude that anomalous metallic phase is intimately associated with inherent inhomogeneities, exhibiting robust behavior at attainable temperatures for strongly granular two-dimensional systems.
\end{abstract}

\maketitle

\section*{Introduction}

An increasing number of recent experiments have been pointing to the possibility of zero temperature transition from a superconducting state to an  ``anomalous metallic regime'' with  $T\to 0$ electronic properties that cannot be understood on the basis of conventional Fermi liquid/Drude theory (for a recent review see \cite{KKS2019}).  In particular it has been argued that the anomalous metal  behaves as a ``failed superconductor,'' a  state in which there are significant  superconducting correlations, yet the system fails to globally condense even as $T\to 0$, settling at a finite conductivity that can be orders of magnitude larger than the Drude conductivity. Among its striking features, current  in the anomalous metal regime is carried by bosonic quantum fluctuations of the superconducting order parameter, exhibiting giant positive magneto-resistance, a much suppressed Hall response \cite{Breznay2017}, and absence of cyclotron resonance \cite{Wang2018}. 

Anomalous metallic phases often emerge in searches for a zero-temperature superconductor-insulator transition (SIT) in disordered superconducting films, yielding instead a quantum superconductor-to-metal transition (QSMT), typically triggered by varying external parameters such as magnetic field, gate voltage, and  degree of disorder \cite{Yazdani,Ephron1996,MasonKapitulnik1,MasonKapitulnik2,Qin2006,Eley2013,Saito2015,Bottcher2018,Chen2018,Breznay2017}. Focusing on the magnetic-field tuned SIT (H-SIT), an ``avoided'' transition (that is, a higher temperature signature of H-SIT that gives way to resistance saturation at lower temperatures) is often found above the QSMT, accompanied by low-temperature resistance saturation that may persist on both sides of this transition. While much of the current focus is on anomalous metallic phases proximate to a QSMT, an anomalous metallic regime has also been ubiquitously identified on the insulating side of a putative SIT. Such a behavior was identified for example in amorphous superconductors tuned by magnetic field \cite{MasonKapitulnik1}, or disorder \cite{Couedo2016}, and assumed to be originating from superconducting fluctuations that persist to the high resistance state. This scenario can be rationalized in the presence of inherent inhomogeneities, where a study of the effect of a ground plane next to the sample further suggested the importance of local superconducting phase coherence \cite{Mason2002}.

From the very nature of the phenomenon, it is clear that the failed superconductor is extremely fragile, primarily due to the inhomogeneous nature of the superconducting state. This can be a result of a granular morphology, or a result of microscopic disorder that is ``amplified'' in the superconducting state to yield an effective inhomogeneous microstructure \cite{Shimshoni1998,Ghosal1998,Dubi2007} or unstable nonequilibrium state due to fluctuations \cite{Aronov1980}. Indeed, recent experiments probed the stability of magnetic-field tuned superconducting amorphous indium-oxide \cite{Tamir2019} and MoGe \cite{Dutta2019} films, concluding that an observed metallic state can be largely eliminated by adequately filtering external radiation. Since in practice any such statement is subject to the limited sensitivity of experiment, the ultimate question arises whether in the absence of any external perturbation, the film's resistance will saturate to a finite resistance, may it be smaller than the experimental limit, as $T\to 0$. This challenging question arises following many experiments where attempts to eliminate the metallic phase failed, and by theoretical solution of a model of superconducting grains embedded in a metallic matrix, where such anomalous metallic behavior can occur in the neighborhood of a QSMT \cite{KKS2019}. Thus, it is essential to distinguish a ``fragile superconductor,'' which is sensitive to external perturbations, from an anomalous metal, i.e. a ``failed superconductor'' in which quantum phase fluctuations preclude superconducting long-range-order even as $T\to 0$.

With the reasonable assumption that a superconducting transition in 2D disordered metallic films is dominated by phase fluctuations \cite{LarkinFeigelman,Hruska,OretoKivSp}, and thus can be modeled as superconducting grains embedded in a metallic matrix, we may search for a system where phase fluctuations are enhanced, and material parameters can be tuned to allow interrogation of an observed anomalous metallic phase within the sensitivity of the experimental system. Since quantum fluctuations of the phase of an isolated superconducting grain are associated with the charging energy, which is further controlled by the dielectric response of the surrounding matrix, a properly designed granular system should be our starting point.

In this paper we examine the robustness of an observed metallic phase in 2D InOx/In composite system, where a thin layer of amorphous indium-oxide (InOx), tuned to be ``barely metallic'' and non-superconducting, provides the Josephson-coupling between indium (In) islands grown on top of it. The unique microstructure of the composite allows for fine control of the fragility of the zero-field superconducting state. Furthermore, the relatively large In grains ensure thermal equilibrium, and thus global superconductivity is achieved uniquely via phase coherence. This system thus yields a highly tunable yet robust anomalous metallic state. While this metallic phase is observed under careful electromagnetic filtering and within a wide range of linear response, exposing it to external broadband noise or microwave radiation is shown to strengthen the tendency of superconductivity. Furthermore, in a wide range of parameters, the external radiation restores a true superconducting phase (within the sensitivity of the measurements,) which is a direct consequence of enhanced phase coherence. 

Close to the QSMT in the metallic phase, resistance saturates to a value that is much smaller than the normal state ``Drude value,'' and depends on magnetic field as a power law. Increasing the magnetic field, a logarithmic divergence of the conductance is observed, with isotherms merge into an ``avoided'' H-SIT (that is, a higher temperature signature of H-SIT that gives way to resistance saturation as the temperature is lowered). Further increase of the magnetic field yields an anomalous logarithmic divergence of the resistance with a large coefficient \cite{Steiner2005PRL}, which mimics the conductance behavior at lower magnetic fields, and thus suggests some form of duality between the low-field and high-field sides of the crossing point. The origin of the logarithmic behavior can be attributed to the granular nature of the system \cite{Beloborodov2007}.  

Finally we note that this versatile system was recently shown to undergo a true magnetic field tuned SIT (H-SIT) \cite{Hen2020}, similar to that of uniform InOx, exhibiting a ``giant'' magneto-resistance above the H-SIT \cite{Sambandamurthy2004,Steiner2005} and critical behavior that manifests the duality between Cooper pairs and vortices \cite{Breznay2016}. This in turn will be a key starting point to our discussion below on ``avoided'' H-SIT.

\begin{figure}[ht]
	\centering
	\includegraphics[width=\columnwidth]{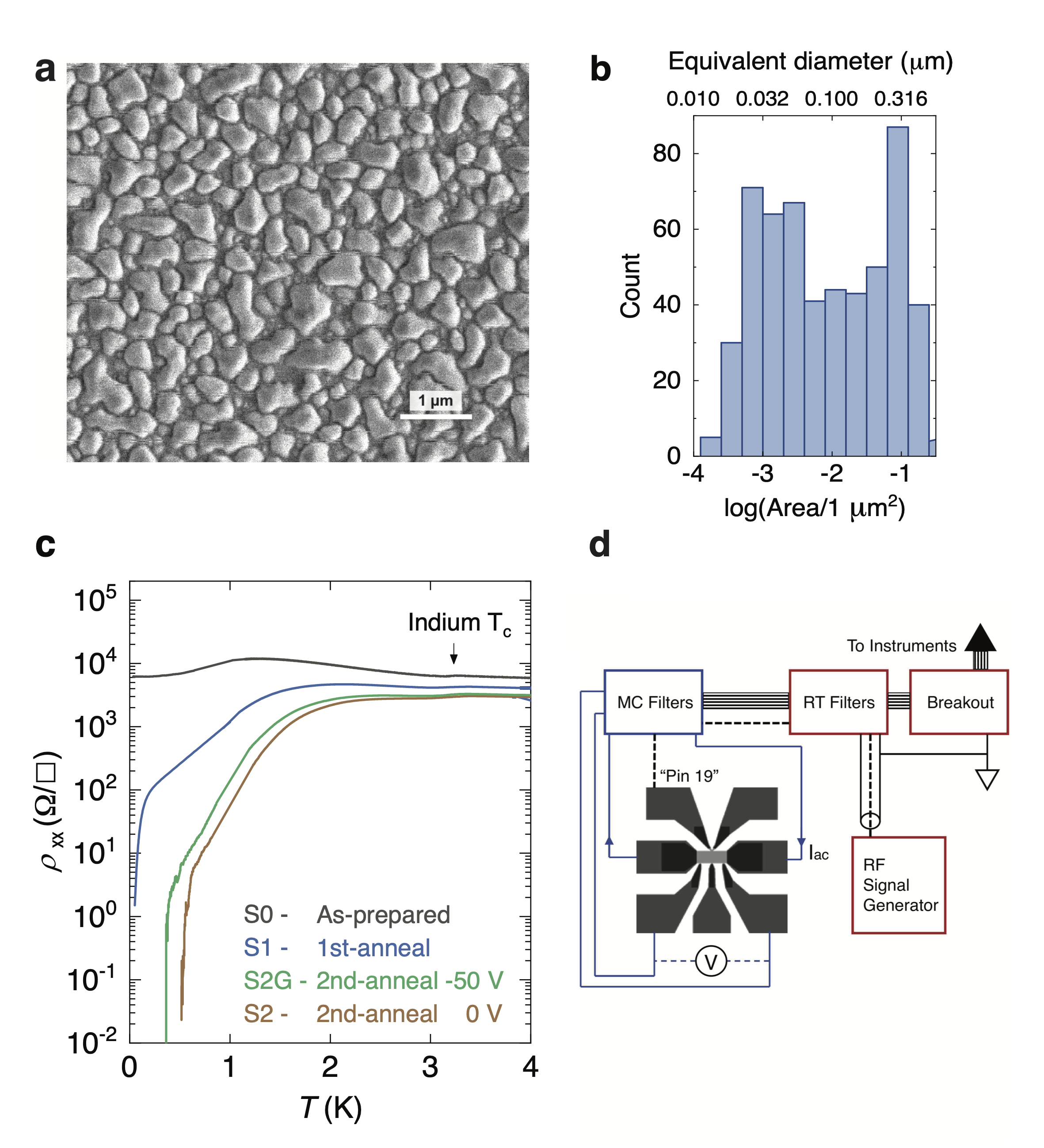}
	\caption{\textbf{Sample characterization.} \textbf{a} Scanning electron microscopy (SEM) micrograph of the InOx/In sample. Bright grains on the foreground are metallic indium, while dark background that lies uniformly underneath is weakly-insulating amorphous InOx. Note the existence of In grains of all scales including the interstitial ones. \textbf{b} Histogram of logarithm of grain area, extracted from the SEM data using a clustering algorithm (see Supplementary Figure 1), shows a broad distribution of grain size. The top axis shows equivalent diameter of a grain. Limitation of image resolution cuts off the distribution below $\sim$0.01\ $\mu$m. \textbf{c} Schematics showing resistance measurement and coupling of RF signal to the sample. See Methods for filter specifications. The RF signal generator generates at output power 0\ dBm (1\ mW) unless otherwise noted. ``Breakout'' is a room-temperature breakout box for measurement wires, where the shielding serves as a common ground for the entire measurement system. \textbf{d} Zero-field resistivity as a function of temperature. Data have been smoothed. Superconducting transition temperature ($T_c\approx3.4\ K$) of indium is indicated by the arrow where resistivity drops by around 10\%. As sample anneals, resistivity decreases and a global superconducting ground state emerges. }
	\label{fig1}
\end{figure}

\section*{Results and Discussions}

As-grown InOx/In samples were prepared intentionally to be initially non-superconducting. Subsequent room-temperature annealing in vacuum with minimal air contact reduced resistivity as demonstrated for 4 sample anneal stages in Fig.~1(c): S0 - as prepared sample, S1 - the sample after 1st anneal, S2 - the sample after a subsequent anneal, and S2G - the second-anneal sample with a -50\ V back gate applied. For every anneal stage including the back-gated, sample resistivity increases when cooled from 300\ K to 3.2\ K similar to plain InOx. Around 3.2\ K, however, resistivity drops by $\sim$ 10\% marking the onset of superconductivity in In grains. Upon further cooling, S0 resistivity fails to exhibit global superconductivity, but instead saturates as $T\rightarrow0$\cite{note_saturated_rho}. As the sample anneals, zero-resistance is almost achieved at base temperature for S1, and can be clearly identified at 0.5\ K and 0.35\ K for S2 and S2G respectively. Hall measurements on the annealed samples confirmed a high carrier density of $4\times10^{15}\ \text{cm}^{-2}$. This value, which is higher but close to that of the pure InOx, is what we expect for a composite system where the non-percolating component exhibits much lower resistance and much higher carrier density. (See supplementary for an estimation of effective carrier density.) Thus, while the -50\ V back gating effectively reduced carrier concentration leading to a reduction in $T_c$, it was not enough to tune S2 through the QSMT. We henceforth focus on S2 and S2G to study the transition between a superconducting and its proximate ground states in a perpendicular magnetic field. 

\begin{figure*}[ht]
	\centering
	\includegraphics[width=2\columnwidth]{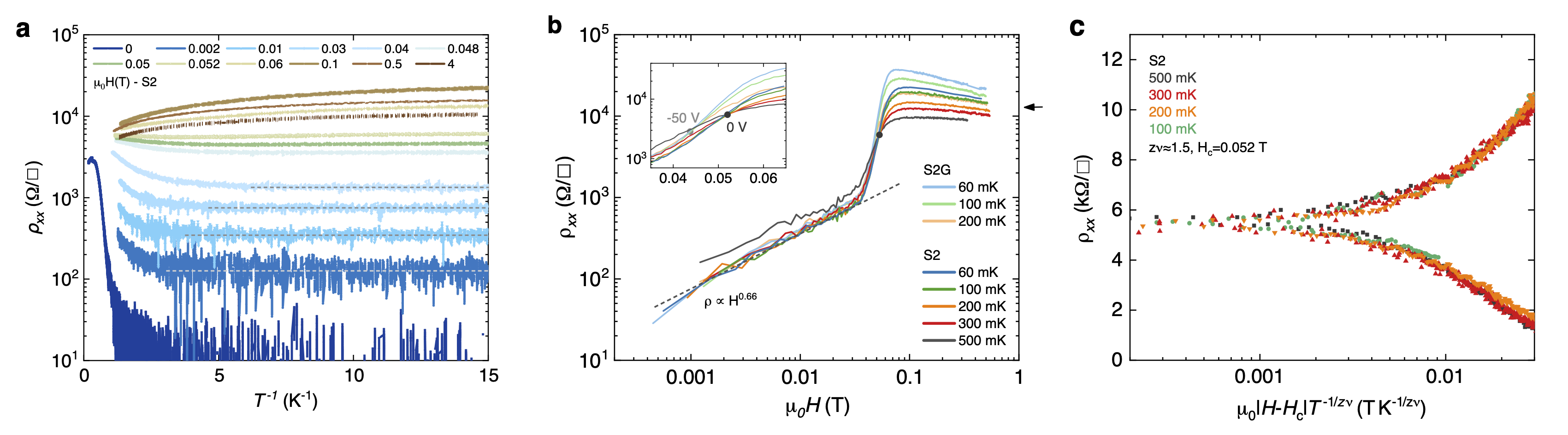}
	\caption{\textbf{Anomalous metallic phase and the avoided H-SIT.} \textbf{a} Arrhenius plot of resistivity in various magnetic fields for sample S2. Robust resistivity saturation can be found in low magnetic field at as high as 300\ mK, indicated by dashed horizontal lines. \textbf{b} Log-log plot of magneto-resistance (MR) at different temperatures for S2 and S2G. In anomalous metallic phase between 0 and 0.04\ T, resistivity is essentially temperature-independent below 300\ mK. The data collapse into a single power law $\rho\propto H^{0.66}$, indicated by the dashed line. The arrow indicates resistivity of S2 at 4\ T. (Inset) Expanded view of the same data between 0.035\ T and 0.065\ T shows isotherm crossing points separating metallic from insulating behavior. \textbf{c} Scaling function for resistivity $\rho(|H-H_c|T^{-1/z\nu})$ of S2 at 100, 200, 300, and 500\ mK. The best data collapse occurs for $z\nu\approx1.5$.}
	\label{fig2}
\end{figure*}

\subsection*{General trends}

With the application of perpendicular magnetic field, the zero-resistance superconducting state in the In/InOx system is rapidly terminated at a magnetic-field-tuned QSMT, and an anomalous metallic phase is identified for $0\lesssim \mu_0 H<0.04$\ T. Fig.~2(a) shows resistivity that saturates as $T\rightarrow0$ at a level orders of magnitude lower than the Drude value. As magnetic field increases, resistivity initially rises as a power-law that is essentially temperature-independent below 300\ mK for both S2 and S2G (see Fig.~2(b)). Then, at $H\approx0.04$\ T, resistivity quickly picks up, leading to a giant magneto-resistance (MR) peak. The transition between saturating and diverging temperature dependence of resistivity is marked in the inset of Fig.~2(b) at 0.052\ T and 0.043\ T, for S2 and S2G respectively. This is reminiscent of a true H-SIT, with a transition to boson-dominated insulating ground state \cite{Sambandamurthy2004,Steiner2005}. At $H\approx0.08$\ T, resistivity peaks at $40\ k\Omega/\square$ at the highest. Beyond the peak, resistivity slowly decreases up to 8\ T, but remains orders of magnitude larger than the normal state value, indicating the persistence of superconducting correlation well into the insulating state \cite{Steiner2005,Baturina2007}. Qualitatively, S0 and S1 behave similarly as shown in Supplementary Figure 2, but have much higher resistivity up to $0.5 M\Omega/\square$ at the peak.

\subsection*{Low-field regime: a failed superconductor}

Starting with the low-field regime, where a failed superconducting phase yields a saturated resistance, the temperature-independent power-law MR before the rapid upturn, $\rho(H)\propto H^{0.66}$, fully overlaps with the anomalous metallic regime. Exponentiating this behavior, this is equivalent to a quantum-tunneling expression with $\rho(H)\propto\exp(-U(H)/k_B T_{\text{eff}})$, previously observed in amorphous MoGe films \cite{Ephron1996}, where the logarithmic field dependence of $U(H)$ was  attributed to activation of free dislocations in a vortex lattice with short spatial correlations \cite{Feigelman1990}. Since this mechanism should strongly depend on the morphology of the film, it is no surprise that the magneto-resistance curves in the anomalous metallic phase for both S2 and S2G films collapse on the same curve in Fig.~2(b). Slight differences between the two samples are observed as the resistance increases above saturation towards a putative crossing point, which we will dub as ``avoided true H-SIT.''  We will come back to analyze the form of the resistance increase.

\subsection*{Avoided H-SIT}

Beyond the resistivity upturn at $H\approx 0.04\ T$, isotherm crossing points that separate positive MR on the low-field regime from positive MR on the high-field regime are observed (marked by full circles) in the inset of Fig.~2(b). This type of behavior would describe a true H-SIT if indeed a true superconducting phase would be attained, accompanied by divergence of the resistance on the insulating side. Since resistance saturation is observed in the low temperature limit of both regimes, we may consider this avoided criticality as a crossover and attempt scaling at higher temperatures (here $T>100$ mK), before the system is dominated by the anomalous metallic phase. Similar approach was demonstrated previously  \cite{MasonKapitulnik1,chakravartykapitulniandme} to lead to scaling exponents of order $z\nu\approx 1.5$ when we scale the data according to $\rho(T,H)=\rho_c\mathcal{F}(|H-H_c|T^{-1/z\nu})$ in Fig.~2(c). It is also interesting to note that such a behavior with a similar critical exponent was previously observed in a 2D system of aluminum islands coupled by 2DEG \cite{Bottcher2018}, thus can be added to the list of common features between systems of induced superconducting granularity in otherwise presumed homogeneous materials such as amorphous-MoGe \cite{Ephron1996,MasonKapitulnik1} and systems with extreme granularity in ordered \cite{Bottcher2018} and disordered (this manuscript) films.

\subsection*{Behaviors proximate to the ``avoided H-SIT''}

Examination of the resistance trends on both sides of the crossing point reveals an intriguing behavior which may indicate the existence of vestiges of duality despite it being an ``avoided transition.''

Starting with the ``insulating regime,'' again S2 and S2G show similar behavior, exhibiting anomalous logarithmic divergence of resistivity that is much weaker than the commonly observed activation or variable-range hopping trends. As shown in Figs.~3(a) and 3(b), where resistivity is plotted versus logarithm of temperature, within a broad range of temperature, the data can be fit by a straight line in the form $\Delta R/R(1\text{K})=A\cdot R(1\text{K})\ln(1/T)$, where the slope $A=(8.44\pm0.12)\times10^{-5}\ \Omega^{-1}$ for S2 at 4\ T. Similar magnitude slopes are found for other magnetic fields, and for the different anneal stages. The data and particularly the slopes of the log-$T$ behavior are neither compatible with Kondo-effect behavior, nor are they compatible with weak localization corrections. Similar logarithmic divergence has been observed in the high-field insulating state of underdoped La$_{2-x}$Sr$_x$CuO$_4$ \cite{Ando1995}, and in amorphous InOx \cite{Steiner2005PRL}. Invoking inhomogeneous microstructure for these otherwise considered homogeneous materials, a possible connection between granularity and logarithmic divergence of the temperature-dependent resistivity has been previously discussed (see e.g. Ref.~\cite{Beloborodov2007}). We further note that for magnetic fields close to the crossing point, the logarithmic divergence tends to saturation, while increasing the magnetic field beyond the magnetoresistance peak seems to recover a ``normal'' insulating behavior commensurate with the initial 2D resistivity of the film before superconductivity sets in. Since the high field resistivity is much smaller than the peak in magnetoresistance, we interpret the saturation on the insulating side as an increase in local phase coherence due to remnant superconductivity.

\begin{figure}[ht]
	\centering
	\includegraphics[width=\columnwidth]{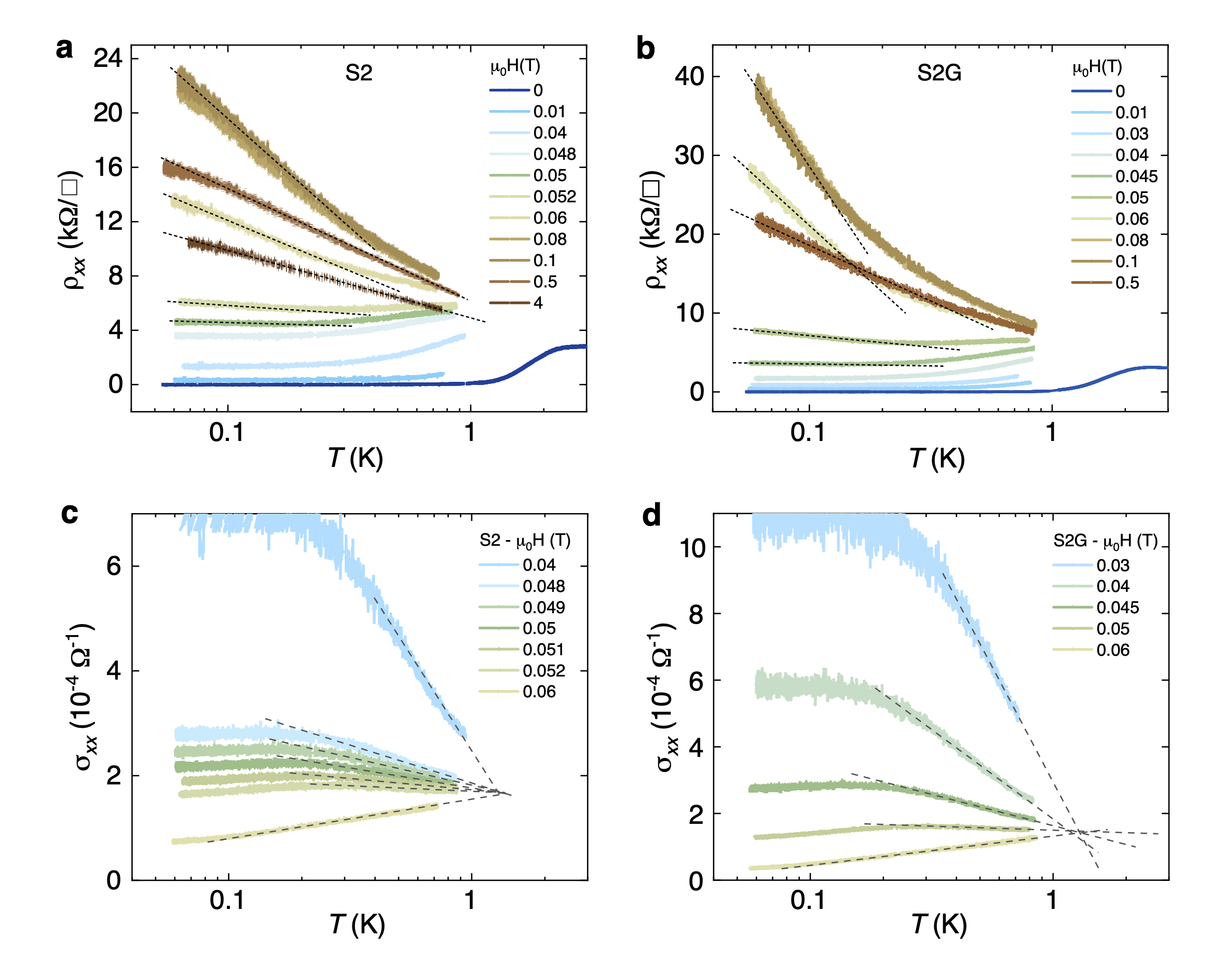}
	\caption{\textbf{Logarithmic divergence of resistivity and conductivity proximate to the avoided H-SIT.} \textbf{a}, \textbf{b} Resistivity versus logarithm of temperature in perpendicular magnetic field, for anneal stage S2 and S2G respectively. Note the difference in resistivity scales. Dotted lines are guidance to the eye, demonstrating the logarithmic divergence of resistivity over a broad temperature range. \textbf{c}, \textbf{d} Conductivity versus logarithm of temperature for S2 and S2G, obtained by inverting the data in (a) and (b). Logarithm divergence of conductivity is found before the saturation sets in at low temperature.}
	\label{fig3}
\end{figure}

Turning to the transition to the ``anomalous metal regime'' below the putative crossing point, we find an anomalous logarithmic divergence of the conductivity (calculated as the inverse resistivity since the Hall contribution is negligible), that is much weaker than the commonly observed activation or variable-range hopping commonly found where a true H-SIT is observed \cite{Breznay2017,Hen2020}. As shown in Figs.~3(c) and 3(d), where conductivity is plotted versus logarithm of temperature, within a broad range of temperature, the data can be fit by a straight line. Similar to the insulating regime, this logarithmic divergence exhibits large prefactor, which again deems it as a crossover to the exponential divergence rather than a weak correction to a normal state conductivity.

Fig.~3 is a striking demonstration of vestiges of the H-SIT quantum critical point in this ``avoided transition.'' It further makes the connection between the resistance saturation on the superconducting side and that on the insulating side, where in our highly granular films it can be attributed to vestiges of local phase coherence in strongly fluctuating superconducting grains.

Finally, we turn to making a more microscopic connection between the logarithmic divergence and the morphology of our films. We will then argue that this connection can be generalized to samples where strong inhomogeneity emerges in the superconducting state following amplification of local disorder \cite{Shimshoni1998,Ghosal1998,Dubi2007}. Focusing on the insulating side, here, with a carefully analyzed microstructure, we are able to show that the observed logarithmic divergence is indeed consistent with originating from the extreme granular inhomogeneity, leading to weak resistivity divergence over an intermediate range of temperature. As shown in Fig.~1(b), the granular structure consists of a hierarchy of grain sizes, spanning over two orders of magnitude, from microscopic interstitial ones to ``large grains'' --- a fraction of a $\mu$m. Hence, in the insulating state, the system consists of a wide range of tunneling barriers between the metallic (or superconducting) islands. These barriers are determined by a competition between normal or Josephson tunneling and charging energy mediated by the dielectric background environment. Assuming a log-uniform distribution of barrier's energy $\Delta$:  $p(\Delta)=1/\Delta$ and summing parallel conductances, we show that the resultant resistance can exhibit weak power law or logarithmic divergence. At much lower temperature, however, the resistivity will diverge much faster in an activated fashion. A more detailed account of our analysis is given in the supplementary. This emerging behavior further supports our intuition that transport is determined by a collective effect of the In grains coupled within the InOx matrix.

\subsection*{Phase diagram}

It was recently demonstrated that where true H-SIT is observed, the quantum phase transition exhibits the emergence of self duality with an exponential diverging variable range hoping behavior of the resistivity above, and the conductivity below, the transition \cite{Breznay2016,Hen2020}. The insulating phase was then identified as a Hall insulator, while the superconducting phase would be an equivalent phase for vortices. Moreover, the assembly of the critical behavior effects were found to be similar to that of quantum Hall to insulator transition \cite{Mulligan2016}, particularly the critical exponents and the emergence of self duality \cite{Breznay2016,Hen2020}. In fact, in attempting to summarize the above results into a phase diagram we observe further similarity between the two phenomena.

\begin{figure}[ht]
	\centering
	\includegraphics[width=0.7\columnwidth]{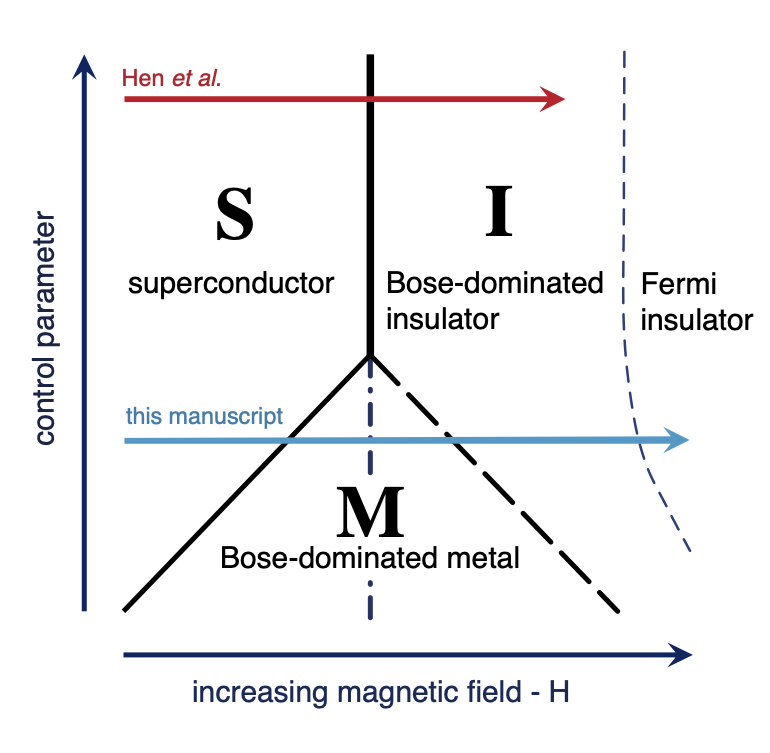} 
	\caption{\textbf{Phase Diagram of Superconductor-Metal-Insulator in 2D.} Phase diagram is drawn as a compilation of the data presented in Hen {\it et al.} \cite{Hen2020} where true H-SIT is observed, together with the results discussed in this manuscript. The thick solid line represents a true H-SIT, with its associated critical behavior and emerging self duality \cite{Breznay2016}. At low field and small control parameter, an  intervening metallic phase emerges from a quantum superconductor-to-metal transition (solid line), showing Bose character until it fades at a crossover to a stronger, Bose-dominated insulating phase (dashed line).  At higher magnetic field pairing is quenched and Fermi-dominated insulating behavior is recovered (thin dashed line). Dash-dotted line represents a line that separates the two regimes of metallic phase, where vestiges of self duality are observed (see text). }
	\label{fig4}
\end{figure}

Fig.~4 is a phase diagram for the In/InOx system based on the experiments of Hen {\it et al.} \cite{Hen2020} observing a true H-SIT and the present data showing an intervening metallic phase. The vertical axis represents an external parameter that controls the nature of the critical behavior. While in uniform films it could be the disorder, in the presence of strong granularity it may represent the distribution of the intergrain couplings. Where material parameters are tuned to observe an anomalous metal (e.g. weak intergrain couplings), a putative crossing point is observed as well with a non-universal critical resistance and different critical exponents, typically $z\nu \approx 1.5$. The fact that the resistance above the crossing resembles the conductance below the crossing, where both lead to a regime of saturated resistance suggests that vestiges of the self-duality line that emerges at the true H-SIT remain also in the regime where a metallic phase emerges, hence marks two dual regimes of anomalous metallicity. This observation is marked with the dash-dotted line in the phase diagram.  Thus, while for true H-SIT a $\sigma_{xx}\sim e^{(\Delta_s/T)^\delta}$ on the superconducting side transforms to $\rho_{xx}\sim e^{(\Delta_I/T)^\delta}$ in the insulating side \cite{Breznay2016}  (here $\Delta_s$ and $\Delta_I$ are activation gap scales for the superconducting and insulating regimes respectively). At weaker control parameter a  $\sigma_{xx}\sim {\rm ln}(\Delta_s/T)$ that appears following the resistance saturation crosses over to $\rho_{xx}\sim {\rm ln}(\Delta_I/T)$ with increasing magnetic field. The insulating side then exhibits further increase of the resistance before pairing is quenched and Fermi-dominated insulating behavior is recovered.

\subsection*{Possible non-equilibrium effects and response to external radiation }

Despite extensive filtering at different stages of our dilution refrigerator, the anomalous metallic phase appears robust. Nonetheless, fragility of the underlying superconducting state can cause this inhomogeneous system to be extremely sensitive to environmental perturbation. In a given magnetic field, barriers associated with collective vortex effects are  established throughout the sample, leading to a saturated resistance  commensurate with that field. However, it is important to explore whether the observed resistance is due to non-equilibrium effects where, for example, electrons fail to thermalize with the lattice due to poor electron-phonon coupling, or external noise heats the electrons, thus obstructing coherence in weaker Josephson couplings and disrupting global superconductivity. We already discussed the unique microstructure of our samples, where the large In grains ensure good thermalization of the local pair amplitudes. Indeed, Fig.~5(a) shows linear response of the resistance measurements in the anomalous metallic regime over a wide range of current bias. Non-linearities are only observed at higher currents, and are presumably associated with junctions locally exceeding their critical current.

\begin{figure}[ht]
	\centering
	\includegraphics[width=0.9\columnwidth]{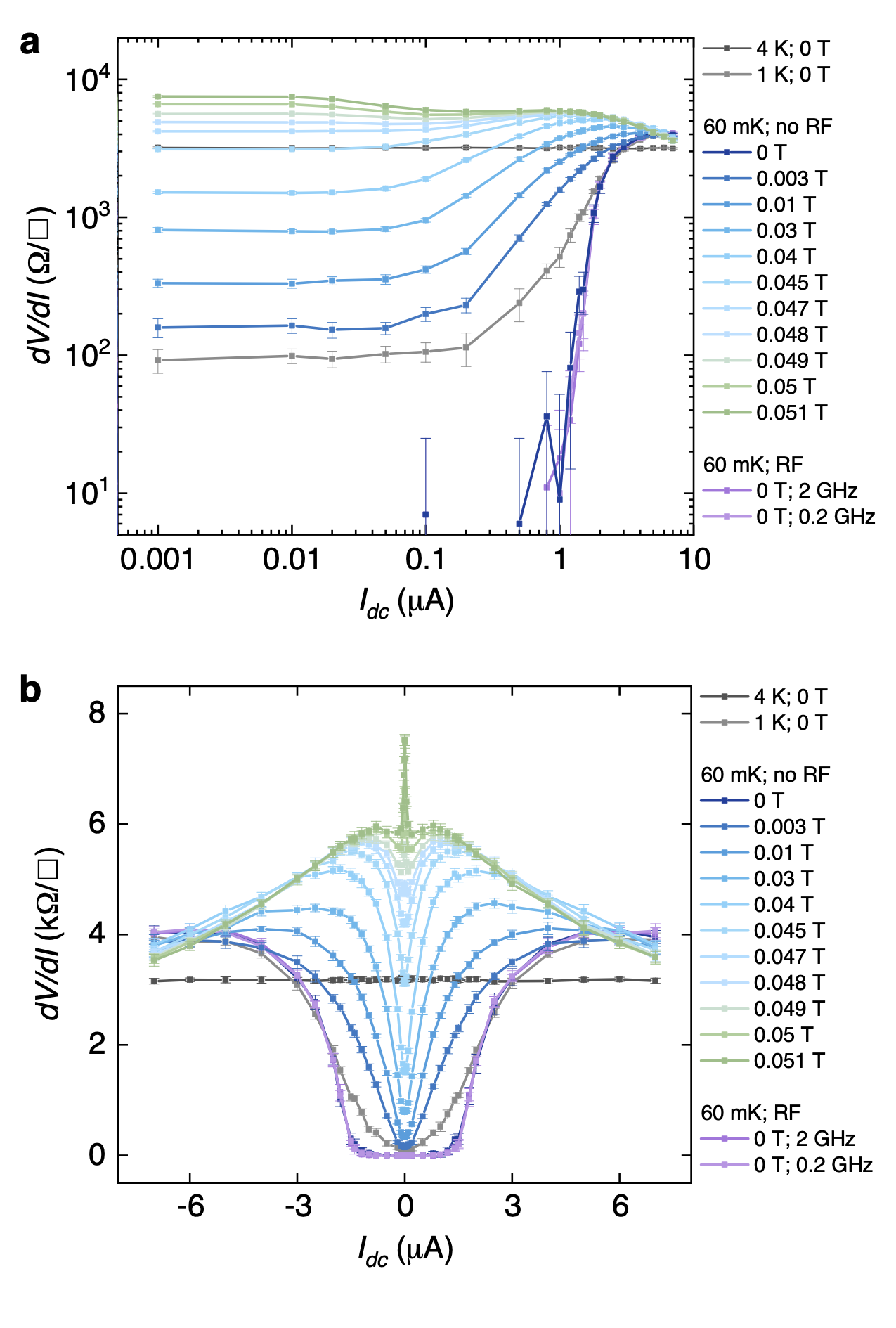}
	\caption{\textbf{Differential resistivity in perpendicular magnetic field or subject to RF injection.} \textbf{a} Differential resistivity $dV/dI$ versus direct current bias for S2G at 60\ mK, showing a superconducting critical field around 1\ $\mu$A. Injection of RF signal of 2\ GHz (dark purple) and 0.2\ GHz (light purple) leads to a slight increase in critical current. Linear response below at least 20\ nA is demonstrated in the anomalous metallic phase. Dark gray curve is $dV/dI$ at 4\ K, marking the normal state value. \textbf{b} Same data shown in full linear scale. Zero-bias features evolve from a dip at low field to a sharp peak at higher field, corresponding to the anomalous metallic phase and insulating phase, respectively. }
	\label{fig5}
\end{figure}

Having established linear response, we further need to explore the possible influence of external radiation on the occurrence of anomalous metallicity. To test for this effect, we may introduce external noise by removing successive layers of filtering in a controlled fashion, or by injecting radio-frequency (RF) signal into the sample and measure its response in resistivity.

\begin{figure}[ht]
	\centering
	\includegraphics[width=\columnwidth]{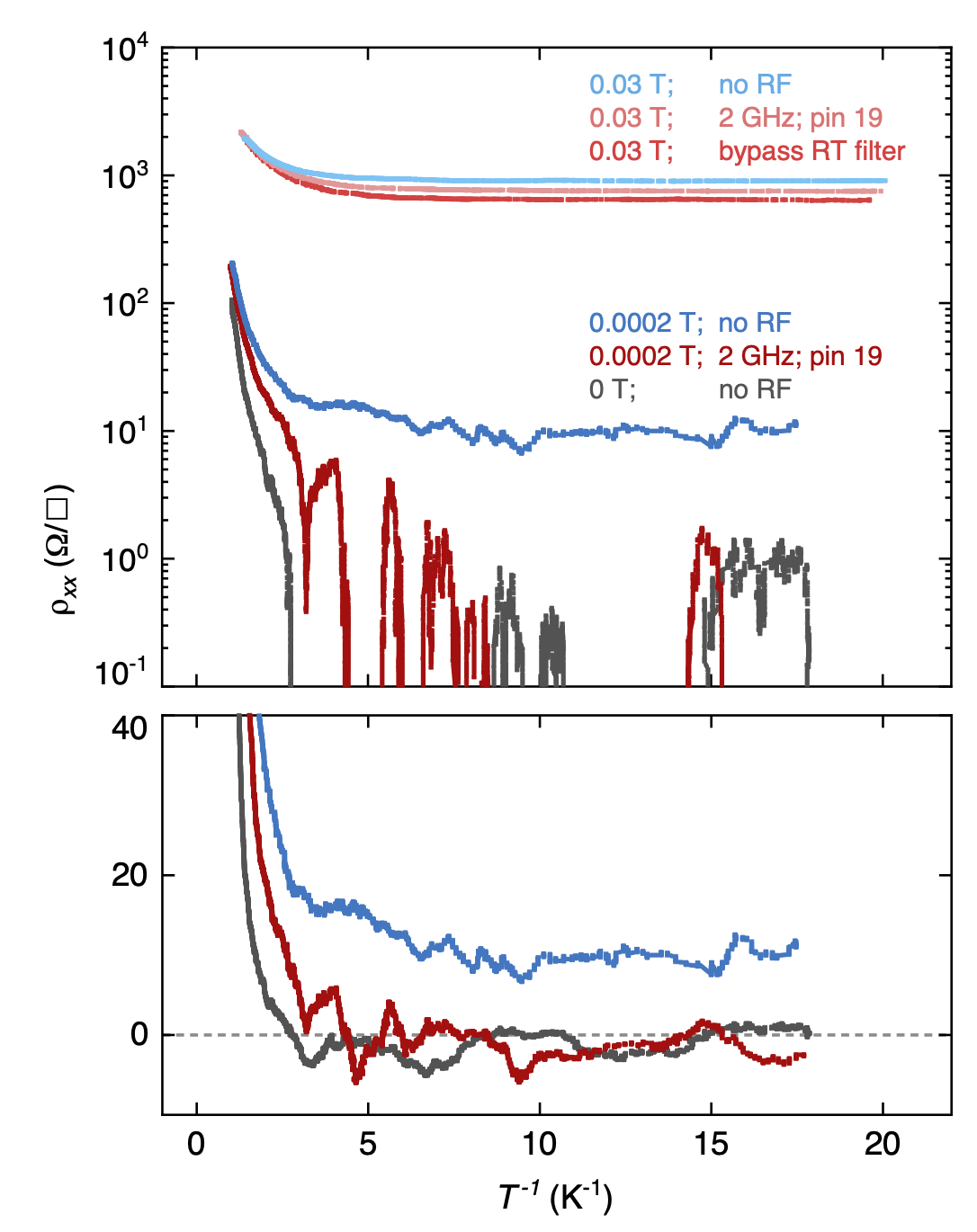}
	\caption{\textbf{Radio-frequency (RF) signal and broad-band noise enhancing tendency towards superconductivity.} Enhanced conductivity by coupling RF signal to the sample S2G, at 0.03\ T and 0.0002\ T. In the latter case, a zero-resistance state emerges as a result of injecting 2\ GHz signal. ``No RF'' means no output power at RF signal generator. ``Pin 19'' denotes a voltage lead connected to the sample in the middle along the Hall bar, see Fig.~1(d). Other wires are checked for de-coupling from the sample at RF, see supplementary. ``Bypass RT filter'' means RT filter is bypassed and broadband noise enters signal line. (No RF signal is injected in this case; data smoothed for higher visibility.) See supplementary for an estimate of RF power transmitted to the sample. Lower panel shows 0.0002\ T data in linear scale, further demonstrating the effect of RF injection.}
\label{fig6}
\end{figure}

We emphasize that our objective is not to study straight forward effects of well coupled radiation to the sample (see discussion below), but rather to inject radiation into various ports that lead to the sample or sample space in order to mimic poor decoupling of the low temperature measurement from the external world. This is important in view of recent studies that suggest that some observations of anomalous metallic state are a consequence of poor filtering and enhanced coupling to the external world.

In Fig.~6, we show Arrhenius plot of resistivity under different experimental conditions. By either introducing 2\ GHz signal or bypassing the room temperature low-pass filter (see Fig.~1(d)), resistivity is lowered by as much as $\sim$ 25\% at 0.03\ T. This is contrary to the electron heating scenario, where an elevated electron temperature would raise saturation resistance. In an extreme case, at a field strength of 0.0002\ T or 2\ Gauss, the anomalous metallic behavior is suppressed completely and true superconductivity is recovered, where resistivity drops below our measurement sensitivity (See the lower panel in Fig.~6). It is interesting to note that in a completely different system of nanopore-modulated YBCO thin films, where anomalous metallic state has been observed, the removal of filtering resulted in a similar reduction of the saturated resistance at low temperatures \cite{Yang2019}. This result was taken as a proof of true cooling of electrons in that highly inhomogeneous system. However, the introduction of deliberate microwave radiation in addition to removal of parts of the filtering system can further illuminate the nature of the robustness of the metallic phase and in particular the effect of re-entrant superconductivity.

Discovered over fifty years ago, microwave-enhanced superconductivity has been studied experimentally in constraint-type \cite{Wyatt1966,Dayem1967} and superconductor-normal-superconductor (SNS) Josephson junctions \cite{Notarys1973,Warlaumont1979}. Dubbed as ``microwave-stimulated superconductivity,'' \cite{Klapwijk2020} this phenomenon was found to be ubiquitous in studies of SNS weak links. Initial theoretical understanding of this phenomenon invoked a non-equilibrium gap-enhancement proposed by Eliashberg \cite{Eliashberg1970}. However, more recent analyses, especially taking into account proximity-effect under external radiation, concluded that much of the enhancement of critical current in such SNS junctions arises from enhanced phase coherence (for a recent review see Klapwijk and de Visser \cite{Klapwijk2020}.)

As demonstrated above, global superconductivity in the the In/InOx system is established via phase coherence among the In grains, coupled through the underlying InOx layer. In the presence of magnetic field, pinned vs. mobile collective vortex effects reflect the restoration or loss of local phase coherence respectively. Thus, a metallic state appears over a zero-resistance superconducting state when a vortex-induced global phase slip appears --- marking the loss of global phase coherence. However, the emerging metallic phase retains much of the character of the superconducting phase, where significant superconducting correlations are present, thus establishing a resistance much lower  than the respective ``Drude resistance.''  The microwave enhanced conductivity of this metallic state, and the restoration of the superconducting state at very low fields are a manifestation of the robustness of this failed superconducting state.

We return to Fig.~5, which shows non-linear differential resistance $dV/dI$ in presence of a direct current bias. In zero magnetic field, superconductivity is quenched by a critical current of 1\ $\mu$A, which is slightly enhanced by application of RF signal. In an external magnetic field, $dV/dI$ yields a zero-bias minimum value in low field, and a zero-bias peak in relatively large field, more clearly shown in full linear scale in Fig.~5(b).

The evolution of the differential resistance behavior displayed in Fig.~5 has been observed previously in a 2D system of aluminum islands, coupled with a gated 2D electron gas \cite{Bottcher2018}. Such a system may be considered an ordered array of islands, as shown by the oscillations in the array's resistance at integer and certain fraction of flux quanta entering their periodic sample. While the overall behavior of the two systems is the same, the random distribution of the In islands in our system blurs any possible ordered oscillation in the magneto-resistance, although it may explain the non-smooth MR data to be discussed in the next section.

\subsection*{Flux effects}
Fig.~7 shows the effect of RF radiation on the magneto-resistance. When no RF signal is present, as a result of the sample's extreme sensitivity to external magnetic field, we cannot clearly identify a lower critical field of a QSMT. Beyond the limitations in measurement sensitivity and magnet field resolution, the main observation is the large fluctuations in MR, up to 30\% of the averaged behavior. Such repeatable fluctuations are likely a consequence of flux commensuration effect allowing flux quanta into the granular structure. While such an effect shows precise periodicity in ordered array of coupled grains, such as in the coupled array of Al islands \cite{Bottcher2018}, or  in studies of square aluminum wire networks in a magnetic field \cite{Wilks1991}, here the area enclosed by loops connecting indium islands through intergrain coupling is random, and the smeared oscillations we observe are a consequence of the mesoscopic nature of the effect, which in turn is also responsible for the observed low critical field. For example, the deep minimum at $\sim$15 G corresponds to a flux quantum fitting in a loop of average diameter $\sim1.3 \mu$m, which in inspecting Figs.~1(a) and 1(b), would correspond to a fundamental loop that includes the larger grains.  Indeed, similar effects have been observed in granular cuprates \cite{Steinmann1988}, inhomogeneous nanowires \cite{Patel2009} and granular lead-film bridges \cite{Wang2009}.

\begin{figure}[ht]
	\centering
	\includegraphics[width=1.0\columnwidth]{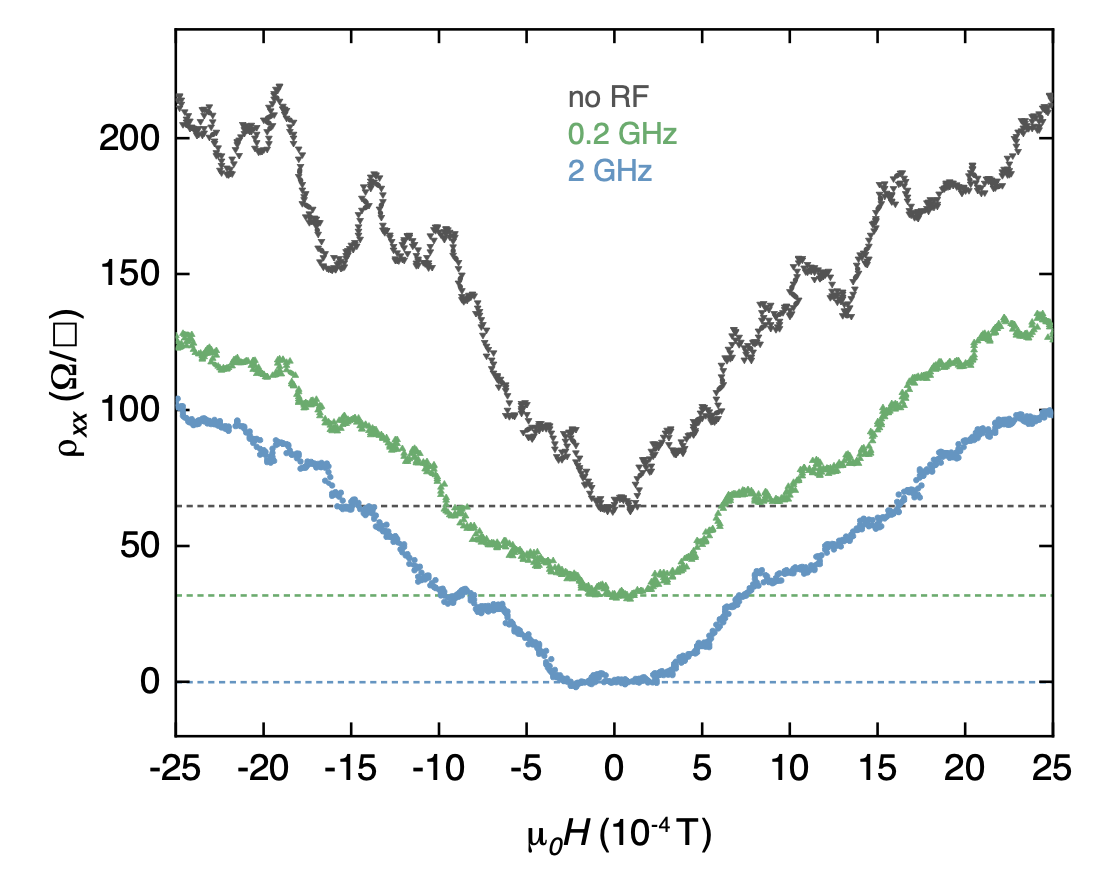} 
	\caption{\textbf{Magneto-resistance of S2G subject to RF signal injection at various frequencies.} Resistivity as a function of magnetic field for S2G at 60\ mK. RF signal of various frequencies is injected into the sample. From top to down: no RF, 0.2\ GHz, and 2\ GHz. $\rho=0$ baseline (dashed-line) of each curve is shifted for clarity.}
	\label{fig7}
\end{figure}

The fact that the MR fluctuations are related to flux penetration through loops of grains connected by junctions is further evidenced from the effect of radiation. As we demonstrated above, the intergrain coupling in the In/InOx/In junctions in our films is enhanced in the presence of microwave radiation, which in turn should make the films effectively more homogeneous and thus reduce the MR fluctuations. Indeed, Fig.~7 shows the suppression of the MR fluctuations, accompanied by enhancement in critical field in the presence of 0.2\ GHz or 2\ GHz RF radiation. 

\section*{Summary}

In this paper we examined the robustness of an observed metallic phase in 2D InOx/In composite system. We found that we can tune this system to a regime of fragile superconductivity that upon application of a weak magnetic field becomes an anomalous metal. While our results are obtained under conditions of careful electromagnetic filtering and within a wide range of linear response, exposure to external broadband noise or microwave radiation is shown to strengthen the tendency of superconductivity, where at low field a global superconducting phase is restored. Increasing the magnetic field exposes an ``avoided H-SIT,'' which above that transition is characterized by an anomalous logarithmic divergence of the resistance, while below the transition a logarithmic divergence of the conductivity is observed. This further highlights the granular nature of the system, and points to possible vestiges of the original duality observed in a true H-SIT. We are then led to conclude that anomalous metallic phase is intimately associated with inherent inhomogeneities, exhibiting robust behavior at attainable temperatures for strongly granular two-dimensional systems.

\section*{Methods}

\subsection*{Sample preparation}
Extreme granular inhomogeneity that features a broad distribution of grain size is achieved by depositing poorly wetting In metal onto uniform amorphous InOx thin film. The underlying layer of 300-\AA\ InOx was electron-beam evaporated onto a commercial lithium-ion conductive glass ceramics (Li$^{+}$-ICGC) substrate (MTI corporation) that enables depletion of electrons up to a 2D carrier density of $\sim10^{14}$ cm$^{-2}$ in back-gating configuration, as was also found in a recent study \cite{Philippi2018}. A deposition rate of 0.32\ \AA/s in an oxygen partial pressure of $9.5\times10^{-6}$\ Torr yielded a weakly insulating film.  Without interrupting the vacuum, at a base pressure of $1\times10^{-7}$\ Torr, indium was evaporated \textit{in situ} at a rate of 5\ \AA/s for 100 seconds, yielding a nominal 500-\AA\ layer of granular indium. The granular structure was confirmed by scanning electron microscopy as shown in Fig.~1(a) along with a histogram showing grain size distribution in Fig.~1(b). The composite film was then patterned in Hall bar geometry (200\ $\mu$m$\times$100\ $\mu$m) using photolithography, before titanium/gold contacts were subsequently patterned onto the sample. Great care was taken to keep the sample below 50 $^\circ$C at all times to preserve the amorphous nature of the  underlying InOx. Effective tuning of disorder, manifested by slow decrease in resistivity, is achieved by annealing the sample in vacuum at room temperature.

\subsection*{Resistivity measurement}

Resistance was measured using standard four-point lock-in technique at 3--13\ Hz using 0.1--1\ nA excitation current. Linear response was verified at various temperatures and magnetic fields. Measurement and filtering schematics can be found in Fig.~1(d). 

\subsection*{RF filtering}

Extensive filtering minimizes electron heating caused by external radiation and improves electron thermalization. Twisted-pair signal lines are filtered at mixing chamber plate using a commercial QFilter RC/RF filter (QDevil ApS), offering over -8\ dB attenuation above 100\ kHz and -50\ dB above 300\ MHz. Additionally, all signal lines are filtered at room temperature by a commercial in-line $\pi$-filter (API technologies corporation), providing an extra -50\ dB attenuation above 200\ MHz. Gate voltage was applied through a room-temperature-filtered twisted-pair well thermally-anchored at mixing chamber plate. Sample phonon temperature is measured by a calibrated on-chip Ruthenium-Oxide (RuO$_2$) thermometer positioned close to the sample.

\section*{Data Availability}

The data that support the findings of this study are available from the corresponding author upon reasonable request.

\section*{Acknowledgments}

We acknowledge discussions with Boris Spivak, Steven Kivelson, Sri Raghu and Teun Klapwijk. Work at Stanford University was supported by the National Science Foundation Grant NSF-DMR-1808385. Work at Tel-Aviv University was supported by the US-Israel Binational Science Foundation (Grant No. 2014098). We thank Sejoon Lim for assistance with SEM. Part of this work was performed at the Stanford Nano Shared Facilities (SNSF), supported by the National Science Foundation under award ECCS-1542152.

\section*{Author Contribution}

X.Z., B.H., A.P., and A.K. designed the experiments. X.Z. grew the samples and performed measurements. X.Z. and A.K. analyzed the data and wrote the manuscript.

\section*{Additional Information}

\subsection*{Supplementary Information}

[To be updated by publisher]

\subsection*{Competing Interests}

The authors declare no competing financial or non-financial interests.

\renewcommand{\thefigure}{S\arabic{figure}}
\renewcommand{\theequation}{S.\arabic{equation}}
\renewcommand{\thetable}{S\arabic{table}}
\setcounter{figure}{0}
\section*{Supplementary Information}

\section{Grain size distribution from clustering analysis on SEM images}
\begin{figure}[ht]
	\centering
	\includegraphics[width=0.9\columnwidth]{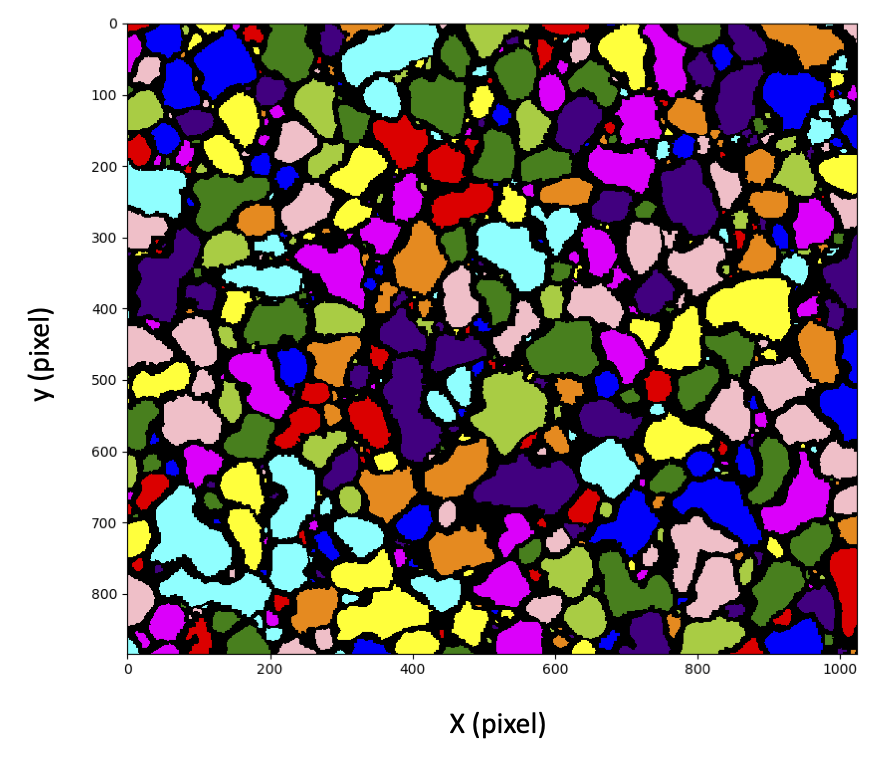} 
	\caption{\textbf{Grain clustering analysis on SEM image.} Identification of grain clusters in the raw image presented in Fig.~1(a) of the main manuscript using open-source scipy and opencv packages in Python. Grain size histogram (Fig.~1(b) of the main manuscript) was calculated based on this result. The axes are shown in a unit of pixel, where 1\ px = 0.006\ $\mu$m. Each color represents a distinct grain.}
	\label{figs1}
\end{figure}
Fig.~\ref{figs1} shows the result of grain clustering analysis on SEM image shown in Fig.~1(a). The raw image was pre-processed using opencv package in Python. To find the contour of grain boundary, we threshold the gray-scale image to create a binary bitmap. A series of morphology transformation (2-dilate/1-erode) was performed to highlight the contours. Then, we took the binary bitmap as a mask, overlaid with the original image, and manually perfected the grain boundary on the mask according to the raw image. After such image processing procedure, we have created a binary bitmap (mask) that outlines correct grain morphology (color: within a grain; black: grain boundary). Subsequently, a standard clustering analysis was performed on the mask and grain statistics was generated by scipy package in Python, rendering the histogram shown in Fig.~1(b). The size distribution of these well-separated grains spans 2 orders of magnitude in length from as large as $\sim$1\ $\mu$m to as small as $\sim$0.01\ $\mu$m (1 or 2 pixels wide), where the distribution is cut-off by limitations in image resolution.

\section{Magnetoresistance of sample S0 and S1}
\begin{figure}[ht]
	\centering
	\subfigure{\label{figs2:a}\includegraphics[width=0.48\columnwidth]{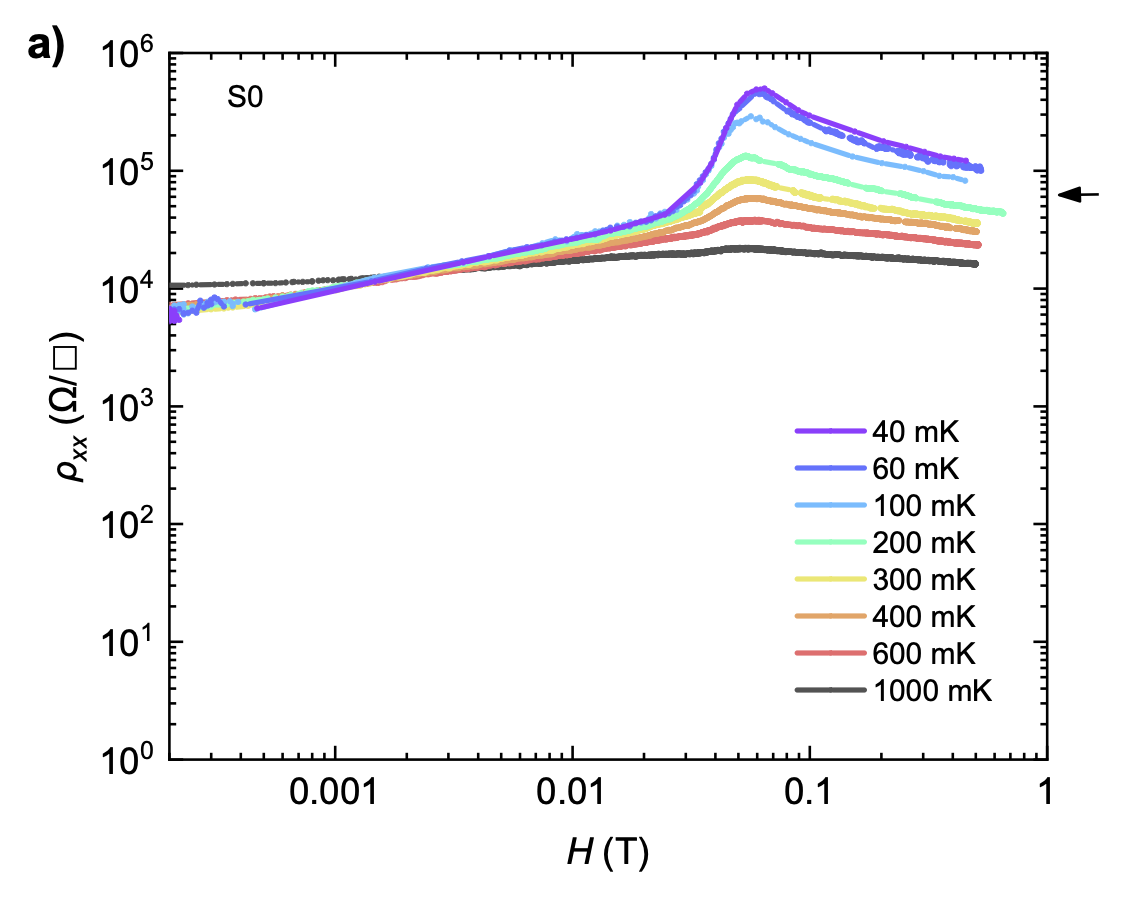}}
	\subfigure{\label{figs2:b}\includegraphics[width=0.48\columnwidth]{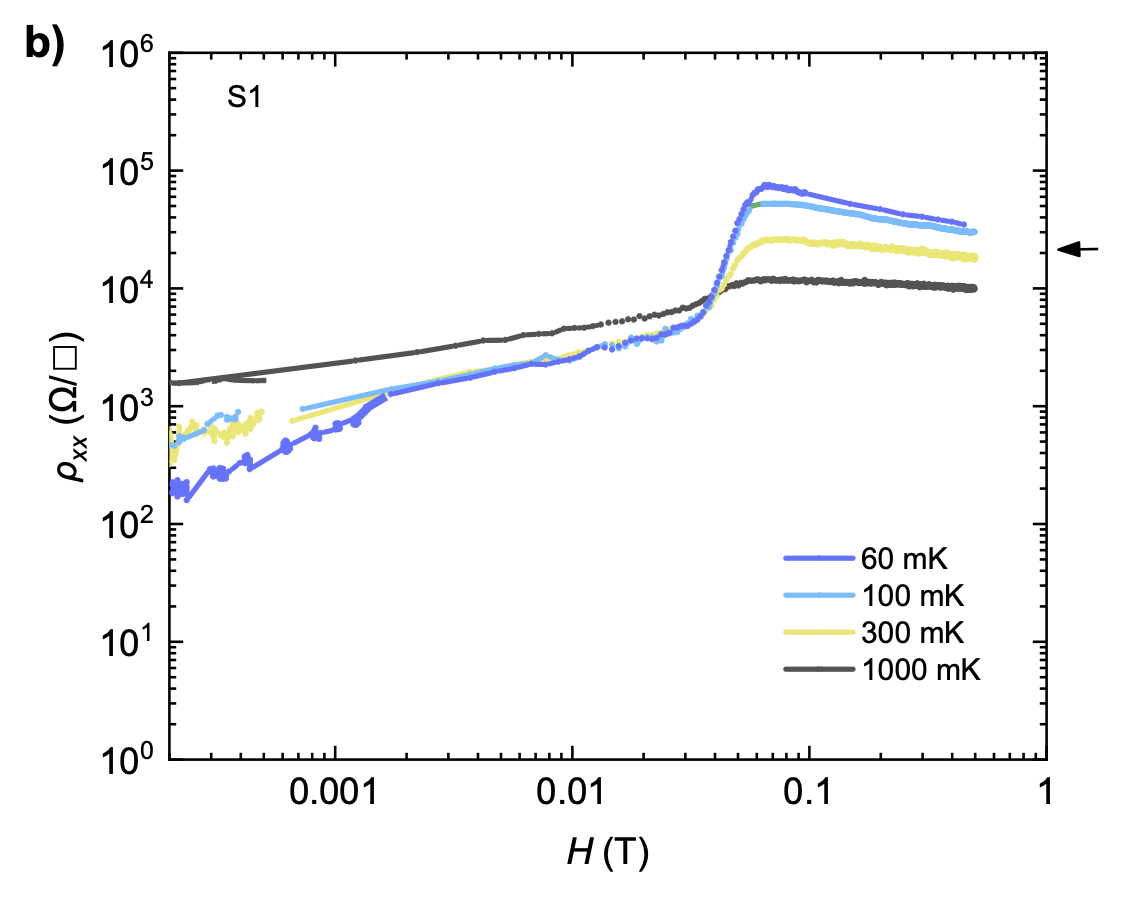}}\\ 
	\subfigure{\label{figs2:c}\includegraphics[width=0.48\columnwidth]{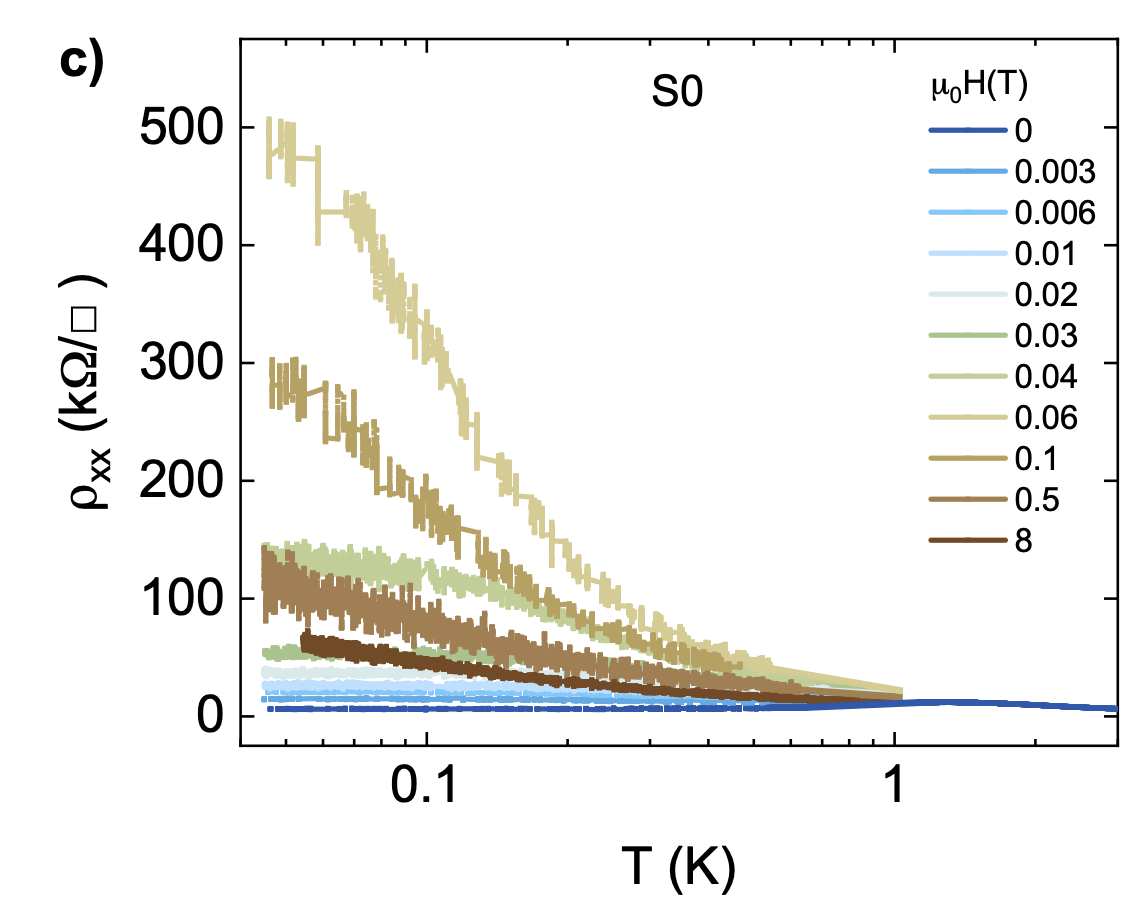}}
	\subfigure{\label{figs2:d}\includegraphics[width=0.48\columnwidth]{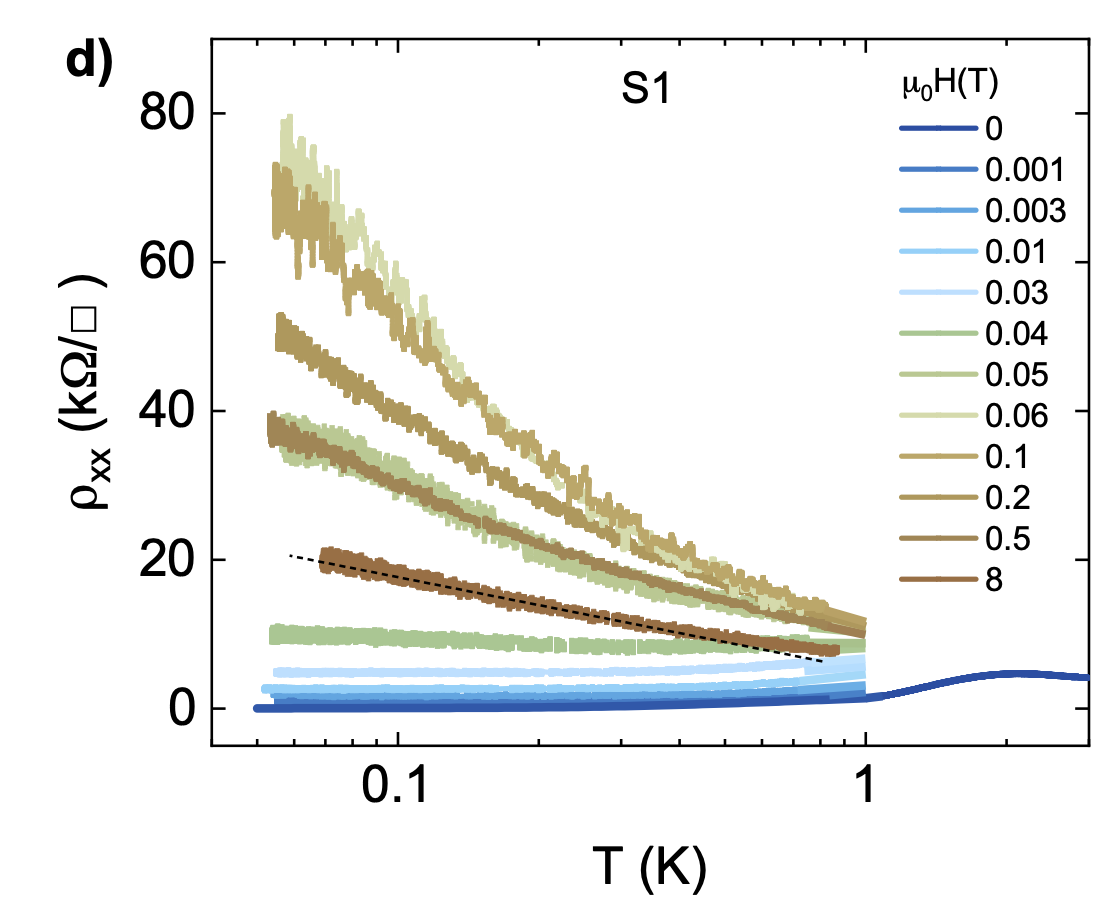}}
	\caption{\textbf{Magneto-resistance and temperature dependence of resistivity of S0 and S1.} \textbf{a}, \textbf{b} Magneto-resistance in log-log plot for S0 and S1 which are not shown in the main text. Similarly, in the saturation regime MR behaves as a power-law. MR peak and the trailing tail are both similar to S2/S2G but at a higher resistivity level. Arrows indicate resistivity at 8\ T. \textbf{c}, \textbf{d} Resistivity versus logarithm of temperature for S0 and S1. Logarithmic divergence similar to those in S2 and S2G can be seen in S1.}
	\label{figs2}
\end{figure}
Fig.~\ref{figs2} shows MR and resistivity of S0 and S1. Note the expanded resistivity scale for these less conductive samples in Fig.~\ref{figs2}a and Fig.~\ref{figs2}b in comparison with Fig.~2(b) of the main manuscript. MR behavior is similar to S2, with an almost temperature-independent power law in low magnetic field. At a magnetic field strength of about 0.04\ T, resistivity quickly picks up, peaking at 0.5\ M$\Omega/\square$ for S0 and 80\ k$\Omega/\square$ for S1. Beyond the peak, MR is negative but remains order of magnitude larger than the normal state value even at 8\ T (indicated by the arrows). Crossing points separating metallic and diverging temperature-dependence are clearly seen in the plots. 

Fig.~\ref{figs2}c and Fig.~\ref{figs2}d shows resistivity of S0 which diverges relatively strongly as $T\to 0$, while for S1 it is similar to S2 and S2G where logarithmic divergence can be found at high field.

\section{Numerical analysis suggesting connection between logarithmic resistivity divergence and extreme granular inhomogeneity}
\begin{figure}[ht]
	\centering
	\includegraphics[width=0.8\columnwidth]{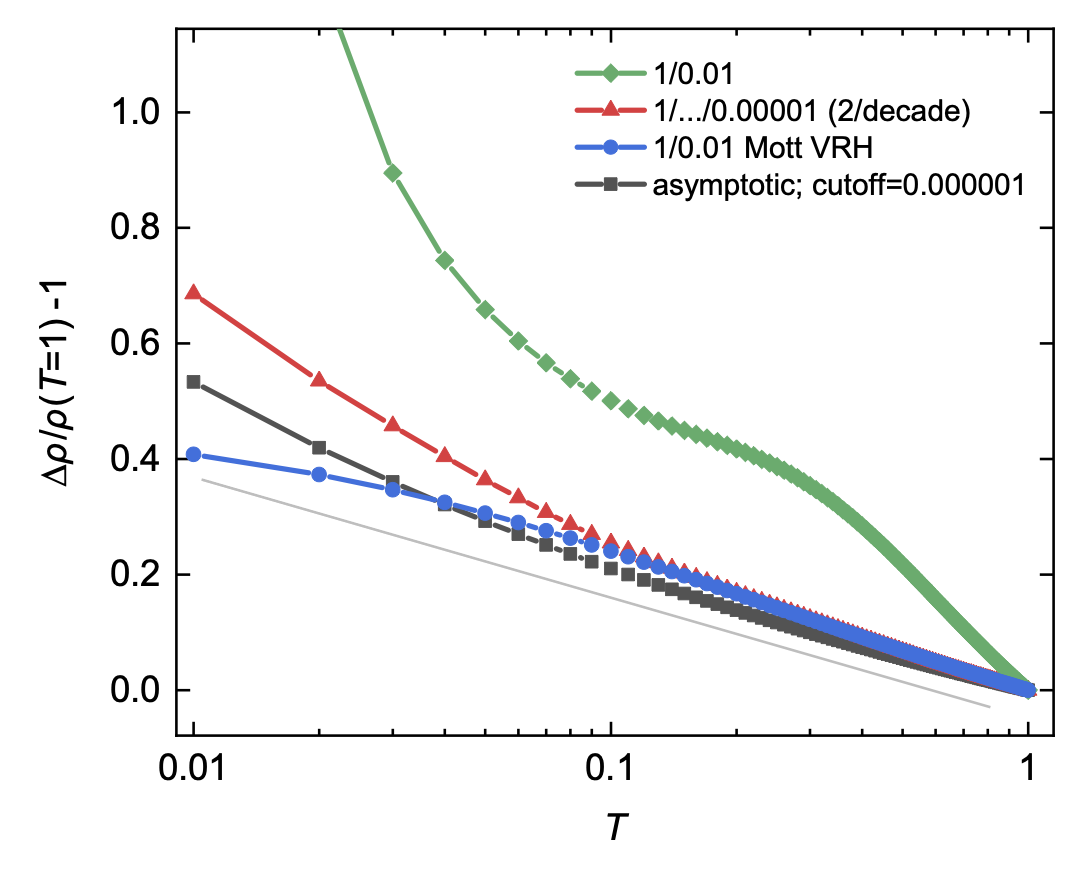} 
	\caption{\textbf{Logarithmic divergence of resistivity calculated numerically.} Resistivity (fractional change) plotted against logarithm of temperature assuming $k_B=1$. For each curve a sum of divergent laws is calculated using specific barrier height indicated in the legends, and then inverted to plot $\rho(T)$. The gray line is a guidance to the eye.}
	\label{figs3}
\end{figure}
Fig.~\ref{figs3} shows the result of our numerical analysis on logarithmic resistivity divergence (or other weak divergence), plotted as fractional change in resistivity compared to resistivity at $T=1$. This analysis is inspired by the experimental result (a weak resistivity divergence when sample is insulating) in Fig.~3 of the main manuscript, and the SEM micrograph (a wide distribution of grains and tunneling barriers.) in Fig.~1(a) of the main manuscript. We consider a parallel conduction model through the wide distribution of tunneling junctions with conductance $\exp(-\Delta/T)$, $\exp(-\delta/T)$, etc. How can the model exhibit a weak divergence? Generally speaking the highest conductance will dominate at low temperatures. However, if we restrict to an intermediate range of temperature, then a combination of parallel conductances through barriers that are orders of magnitude different from each other would allow relatively weaker divergence. Summing all the diagonal terms in a Landauer formula $G\propto\sum \mathcal{T}$, where $\mathcal{T}\propto\exp(-\Delta/T)$ is tunneling transmission through barrier height of $\Delta$. Inverting conductance gives resistance as a function of temperature.
\begin{equation}
	R=G^{-1}\propto(\exp(-\Delta/T)+\dots+\exp(-\delta/T))^{-1}
\end{equation}
The simplest example is $(\exp(-1/T)+\exp(-0.01/T))^{-1}$ (green curve). Blue curve uses VRH law instead of activation ones while red sums eleven activation laws. Black curve is obtained by explicitly calculating the integral discussed in main text using barrier distribution $p(\Delta)=1/\Delta$. The resulting expression for resistivity is the inverse of an exponential integral,
\begin{equation}
	R=(E1(\Delta_\Gamma/T))^{-1}\approx(-\gamma-\ln(\Delta_\Gamma/T)+\Delta_\Gamma/T)^{-1}
\end{equation}
where $\gamma\approx0.577$ is the Euler-Mascheroni constant and $\Delta_\Gamma$ is a lower energy cut-off for barrier height taken to be $\Delta_\Gamma\equiv10^{-6}$.

\section{Ineffective RF injection and analysis of RF transmission}
\begin{figure}[ht]
	\centering
	\includegraphics[width=0.9\columnwidth]{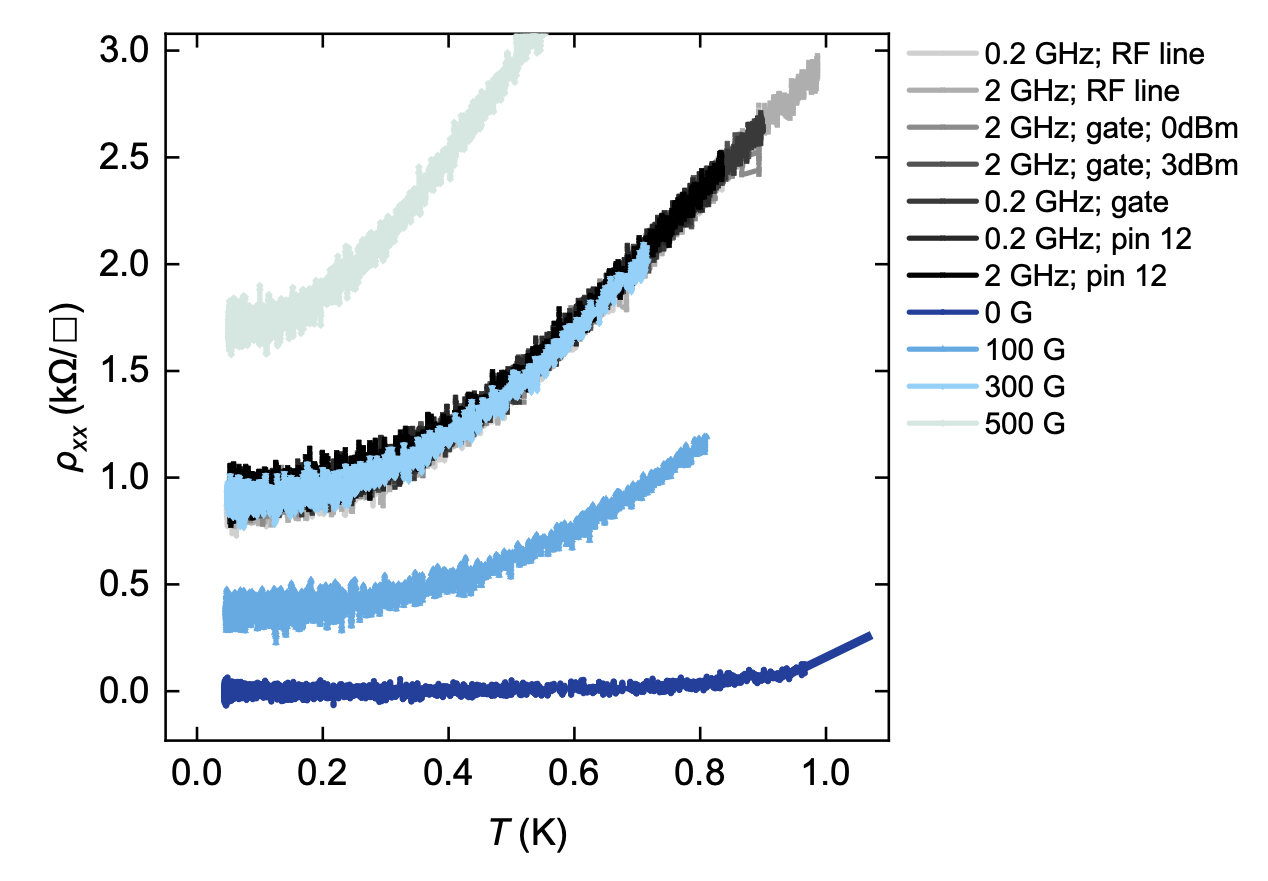} 
	\caption{\textbf{RF injection experiments that have no effect.} Resistivity as a function of temperature in various experimental conditions other than those presented in main text. As the reference group for effective RF injection experiments, compared to no RF injection, no difference in $\rho(T)$ is observed in any of these case.}
	\label{figs4}
\end{figure}

Fig.~4 of the main manuscript shows RF injection experiments that have no effect on sample resistivity. The colored data are reference data with no RF injection, and is the same data as shown in Fig.~3(d). ``RF line'' means a heavily-attenuated coaxial cable that is openly disconnected about 1\ m above the sample inside dilution fridge, not connected to the sample but may be able to broadcast microwave radiation to the sample space. ``Gate'' means the filtered gate wire that is connected to back side of the substrate through gold wire-bond, around 5\ mm away from the sample. ``pin 12'' means a disconnected/grounded wire subject to the same filtering as other twisted-pair measurement wires, located at about 5\ cm above the sample. (1\ cm wavelength corresponds to 30\ GHz in free space.) 

All these wires in principal transmit microwave radiation, such as narrow-band signal generated by RF communication devices (Wi-Fi: 0.9--5.9\ GHz; Cellular: 0.6--2.5\ GHz; AM/FM radio: 1--100\ MHz; the latter two are strongly suppressed in our laboratory), instrument electromagnetic noise, or broad-band black-body radiation from the warmer parts of cryostat/wiring ($\lambda_{peak}=b/T$, where Wien's displacement constant $b\approx2898\ \mu m\cdot K$; 1\ K corresponds to roughly 100\ GHz). 

All RF injection used a minimum output power of 0\ dBm (1\ mW) from a PTS 6400 signal generator. However, the wiring inside our dilution fridge is not impedance matched to the RF signal generator output, and thus the transmitted power is greatly limited. Assuming a 20--40\ dB mismatch loss ($-20\log(1-\Gamma^2)$ in amplitude, where reflection coefficient $\Gamma\approx1-50(\Omega)/Z_L$ and load impedance $Z_L\sim1\ k\Omega$), combined with $\sim$100\ dB (amplitude) filtering at $\sim$1\ GHz and transmission loss along the wires, we expect the power transmitted to the sample is attenuated by $10^{6}$--$10^{7}$ to $\sim$ 100\ pW. The exact power arrived at the sample, however, is not of high importance because we are demonstrating in the main text whether or not an effect takes place. 

\section{RF injection in the insulating regime}

\begin{figure}[ht]
	\centering
	\includegraphics[width=0.8\columnwidth]{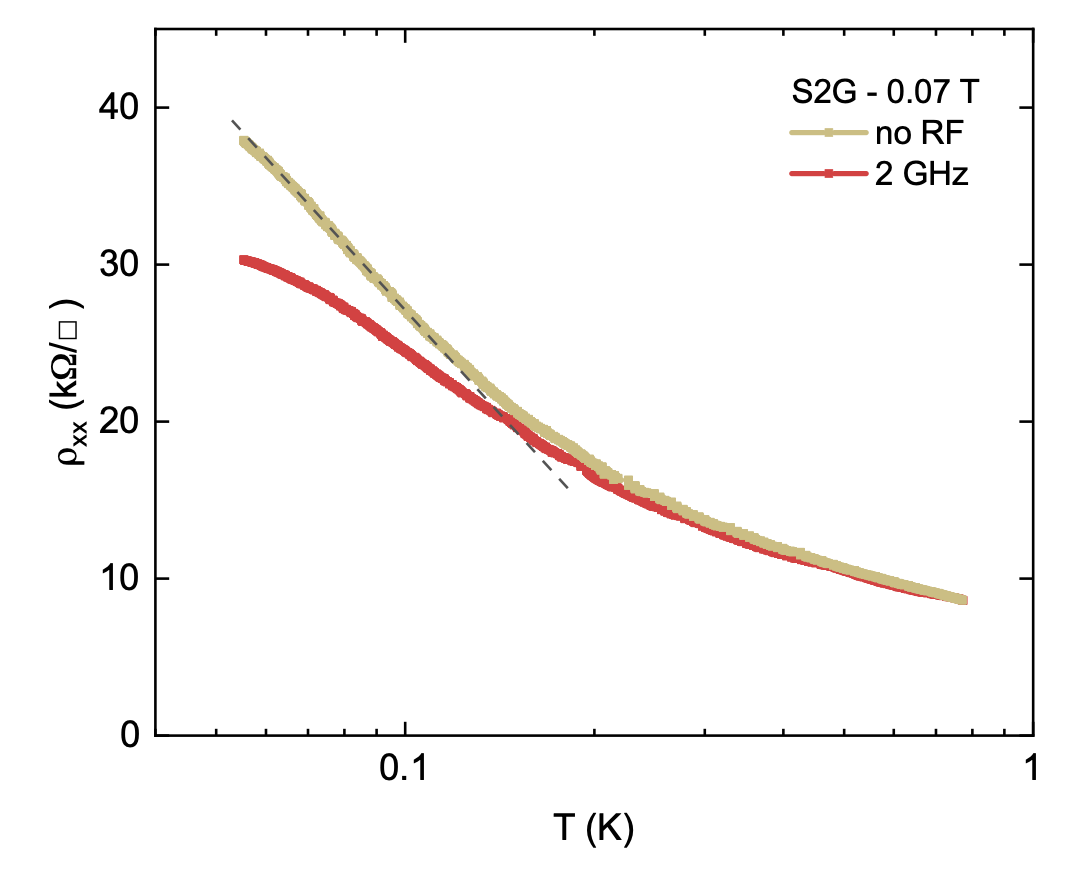} 
	\caption{\textbf{RF injection experiments in the insulating regime.} Resistivity versus logarithm of temperature for S2G at 0.07\ T. The resistivity in the absence of RF signal diverges logarithmically, indicated by the dashed line. Red trace shows effect of 2\ GHz signal coupled to the sample.}
	\label{figs5}
\end{figure}

Fig.~\ref{figs5} shows resistivity versus logarithm of temperature for S2G in insulating regime, where an anomalous logarithmic resistivity divergence is found ubiquitously on the insulating side of the avoided H-SIT. Here we demonstrate that by applying 2\ GHz signal, using the same configuration as in Fig.~6, the resistivity is also reduced by $\sim$20\%. For the same reason discussed in text, such enhancement to the conductivity indicates that the insulating regime is still dominated by local superconductivity.

\section{Carrier density of the InOx/In composite}

Concerning the carrier density of the InOx/In composite, one needs to remember that The indium islands are not continuous while the indium-oxide is uniform and continuous. Therefore, the reasonable model to use to estimate the Hall effect of the composite, from which measurement we extract the carrier density, is by using a two-component composite. We show below that the measured Hall effect and thus extracted carrier density that we obtain is in fact what one would expect for such a composite.\\

The effective Hall conductivity of a two-component inhomogeneous two-dimensional film was previously calculated (see e.g. D. Stroud and D.J. Bergman, Phys. Rev. B 30, 447 (1984).) The exact result in the case where the two components are ``$a$'' and ``$b$'' is:
\begin{equation}
	\frac{\sigma_{xy}^e-\sigma_{xy}^b}{\sigma_{xy}^a-\sigma_{xy}^b}=\frac{[(\sigma_{xx}^e)^2+(\sigma_{xy}^e)^2]-[(\sigma_{xx}^b)^2+(\sigma_{xy}^b)^2]}{[(\sigma_{xx}^a)^2+(\sigma_{xy}^a)^2]-[(\sigma_{xx}^b)^2+(\sigma_{xy}^b)^2]}
	\label{hall}
\end{equation}
Let us take the case where $a$=InOx  and $b$=In. However, the Indium islands remain isolated islands, so component $a$ (InOx) percolates. Now, from our parallel experiments in depositing the various components, the low-temperature 2D resistivity of the InOx is $\sim$10k$\Omega/\Box$, with a carrier density of approx. $1\times10^{20}$ carriers/cm$^3$, while the estimated low temperature resistivity of the pure indium is $\ll 0.01\Omega/\Box$, where we use an averaged thickness of 250 \AA (the nominal deposited thickness of the grains was 500 \AA), and we remember the finite size of the grains limits the indium mean free path, and thus the estimated resistivity is larger than that of uniform indium. Still, it is easy to see that $\sigma_{xx}^b\gg\sigma_{xx}^a$ as well as $\sigma_{xx}^b\gg\sigma_{xx}^e$ since the 2D resistivity of the composite film is $\sim 2.5$k$\Omega/\Box$. The carrier density of the pure indium is $\sim10^{23}$ carriers/cm$^3$.\\

Using the above parameters in equation~\ref{hall}, it is easy to see that if we divide numerator and denominator of the right hand side of equation~\ref{hall} by $(\sigma_{xx}^b)^2$, then to order $(\sigma_{xx}^e/\sigma_{xx}^b)^2$, which is tiny (here we note that the Hall components do not contribute), we obtain (see also Stroud and Bergman paper) that
\begin{equation}
	\sigma_{xy}^e\approx \sigma_{xy}^a
\end{equation}
This is an exact result for the case where the non-percolating ($b$) component is a superconductor with infinite conductivity and zero Hall conductivity.  However, since the pure indium is so much more conductive than the InOx, and its carrier density is so high, which means a very small Hall coefficient, that result approximately holds as we just showed.\\

It is now easy to see that since the InOx ($a$) is the continuous component, the effective Hall conductivity of the composite is that of the InOx.  The Hall resistivity of the component is then:
\begin{align}
	\rho_{xy}^e&=-\frac{\sigma_{xy}^e}{(\sigma_{xx}^e)^2+(\sigma_{xy}^e)^2}\approx -\frac{\sigma_{xy}^a}{(\sigma_{xx}^e)^2+(\sigma_{xy}^e)^2} \\
	&=\frac{\rho_{xy}^a}{[(\rho_{xx}^a)^2+(\rho_{xy}^e)^2 ][(\sigma_{xx}^e)^2+(\sigma_{xy}^e)^2]}
\end{align}
This result leads to the approximation that
\begin{equation}
	\rho_{xy}^e \approx \rho_{xy}^a\left( \frac{\rho_{xx}^e}{\rho_{xx}^a}\right)^2 \approx \frac{1}{16} \rho_{xy}^a
\end{equation}
This means that the carrier density of the composite should be about 16 times larger than that of the pure InOx, which was $\sim1\times10^{20}$. Thus, we expect that the effective carrier density of the component will be $\sim 1.6\times 10^{21}$ carrier/cm$^3$, or, if we multiply by the effective thickness of the layer, we obtain $\sim 9\times10^{15}$carriers/cm$^2$, which is remarkably close to the actual measured number.\\

Finally, we stress that as we write in the title that the composite is two-dimensional, we wish to reiterate that the system is to be thought of as a composite system rather than parallel conduction channels. Indium grains in the composite sets well-defined spatially-inhomogeneous order parameter amplitude, while the underlying indium oxide provides Josephson coupling among indium grains, so the two components are equally important.\\


\begin{thebibliography}{10}
\bibitem{note_saturated_rho}
	\bibinfo{note}{Such samples, which do not show a zero-field global
		superconductivity, will be discussed in a different publication}.
\end{thebibliography}
\end{document}